\documentclass{midl} 

\usepackage{mwe} 

\jmlrproceedings{MIDL 2020}{Medical Imaging with Deep Learning 2020}
\jmlrvolume{}
\jmlryear{}
\jmlrworkshop{MIDL 2020 -- Short Paper}
\editors{}

\title[Overview of Scanner Invariant Representations]{Overview of Scanner Invariant Representations}

\midlauthor{\Name{Daniel Moyer \nametag{$^{1,2,3}$}} \Email{dmoyer@csail.mit.edu}\\
\addr $^{1}$ CSAIL, Massachusetts Institute of Technology, Cambridge, MA, United States\\
\addr $^{2}$ Information Sciences Institute, Marina del Rey, CA, United States \\
\addr $^{3}$ Imaging Genetics Center, Keck School of Medicine, University of Southern California, Los Angeles, CA, United States \AND
\Name{Greg Ver Steeg \nametag{$^{2}$}} \Email{gregv@isi.edu}\\
\Name{Paul M Thompson \nametag{$^{3}$}} \Email{pthomp@usc.edu}
}

\begin{document}

\maketitle

\begin{abstract}
Pooled imaging data from multiple sources is subject to bias from each source. Studies that do not correct for these scanner/site biases at best lose statistical power, and at worst leave spurious correlations in their data. Estimation of the bias effects is non-trivial due to the paucity of data with correspondence across sites, so called "traveling phantom" data, which is expensive to collect. Nevertheless, numerous solutions leveraging direct correspondence have been proposed. In contrast to this, Moyer et al. (2019) proposes an unsupervised solution using invariant representations, one which does not require correspondence and thus does not require paired images. By leveraging the data processing inequality, an invariant representation can then be used to create an image reconstruction that is uninformative of its original source, yet still faithful to the underlying structure. In the present abstract we provide an overview of this method.
\end{abstract}


\section{Introduction and Summary}

In magnetic resonance imaging (MRI), variations in observational conditions, protocol, and equipment induce site-wise and scanner-wise biases in the collected data \cite{chen2014exploration,fortin2017harmonization,jovicich2006reliability}.
Without correcting for these biases, multi-site studies will at best lose statistical power, and in some cases may arrive at erroneous conclusions. It is therefore imperative in multi-site studies to make these corrections; the process of doing so, removing or compensating for unwanted scanner/site-wise variations, is known as harmonization. In the present work we focus on harmonization for diffusion MRI (dMRI), a modality known to have scanner/site biases \cite{correia2009looking,giannelli2010dependence,pagani2010intercenter,papinutto2013reproducibility,vollmar2010identical,white2009global,zhan2010does,zhan2013magnetic,zhan2012spatial} as well as several extra possible degrees of freedom with respect to protocol (e.g., angular resolution, $b$-values, gradient waveform choice, etc.).

Previous work has largely focused on the summary statistic level (e.g. Fractional Anisotropy) \cite{fortin2017harmonization,zavaliangos2018diffusion} or on supervised cases where pairs of images from the same subject collected with different scanners are provided \cite{blumberg2018deeper,tanno2017bayesian}. These methods attempt to estimate the relative effects of each scanner/site, either in derived measures or in the original data domain.

Moyer et al. \cite{moyer2019scanner} present an unsupervised method that instead learns a representation of the images that is uninformed of the scanner/site at which they were collected, yet also one that is otherwise maximally informative of the image. Reconstructions from this uninformed representation will then be uninformed of their original scanner/site context, a result which follows from data processing inequality.

\begin{figure}[t!]
\centering
\includegraphics[width=0.9\textwidth]{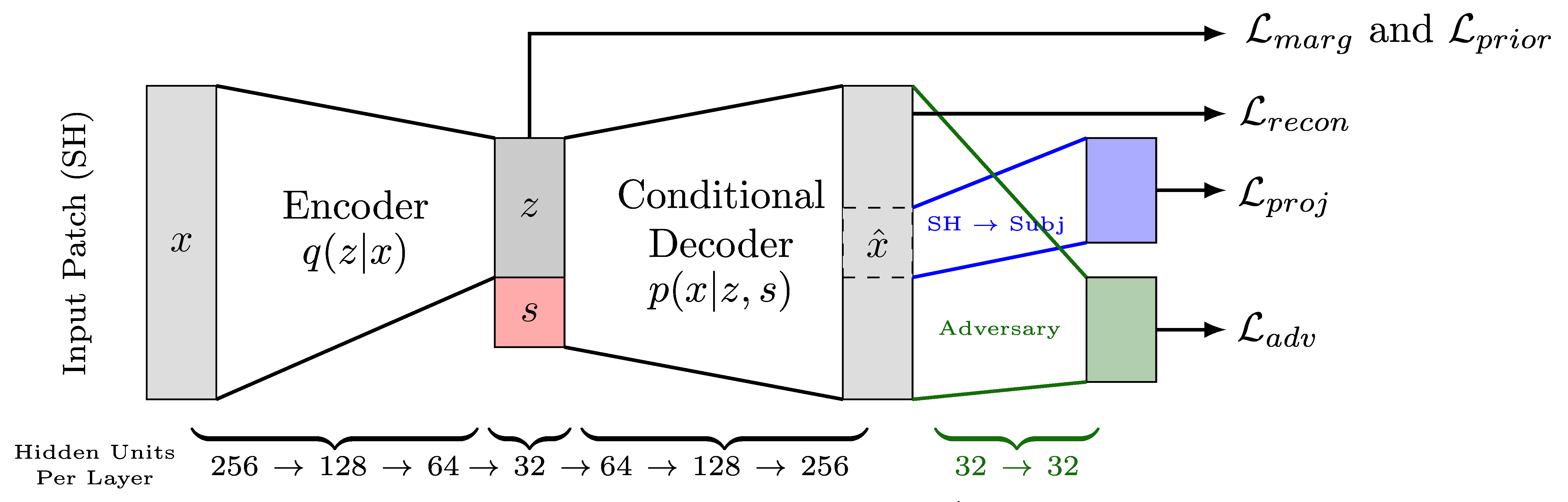}
\caption{Diagram describing network configuration.}
\label{fig:diagram}
\end{figure}

The construction of invariant representations is, generally, non-trivial. A previous paper by Moyer et al. \cite{NIPS2018_8122} show that this can be done by compressed conditional auto-encoder, where the learned encoding becomes uninformed of the conditional factor under compression. This leads to the following procedure:
\begin{enumerate}
\item Construct an auto-encoder for image data, using compressive regularization (e.g. penalizing $I(x,z)$ for data $x$ and encoding $z$), and condition the output on the site.
\item Train the auto-encoder on images from each scanner/site independently.
\item At test time, manipulate the conditional decoder to remap images through the learned invariant code to a single scanner/site context.
\end{enumerate}
As described in \cite{NIPS2018_8122}, conditional architectures such as the one described here (Fig. \ref{fig:diagram}) penalize learned site information $I(z,s)$. In reducing site information in $z$ this architecture reduces $I(\hat{x},z)$, the site information in the image reconstruction.

Due to hardware limitations, the method was applied patch-wise. The authors further included an adversary on the patch-wise output. To generalize representations across particular gradient directions, a spherical harmonics representation was used at the voxel level, which was projected back to subject specific directions when calculating reconstruction loss. The proposed auto-encoders may be learned either for the two-site case (``single-task'') or the multi-site case (``multi-task'').

\section{Overview of the Empirical Evaluation}

Moyer et al. \cite{moyer2019scanner} present an evaluation of this method on the 2018 CDMRI Challenge Dataset \cite{ning2018muti,tax2019cross,taxcross}, which is composed of images from 15 subjects. As described in \cite{ning2018muti} data from each subject were collected on a 3 T GE Excite-HD ``Connectom'' and a 3 T Siemens Prisma scanner, each with two separate protocols. One protocol matches between the scanners at a low resolution (2.4 mm iso., $b={1200,3000}$ with 30 grad. directions), and another which does not match at a high resolution (1.5 mm and 1.2mm iso., same shells 60 grad. directions). This creates 4 ``sites'' (denoted P30,P60,C30,C60). Scans are mapped to and from P30 for each other site.

\begin{figure}[t!]
\footnotesize
\centering
\includegraphics[width=0.9\textwidth]{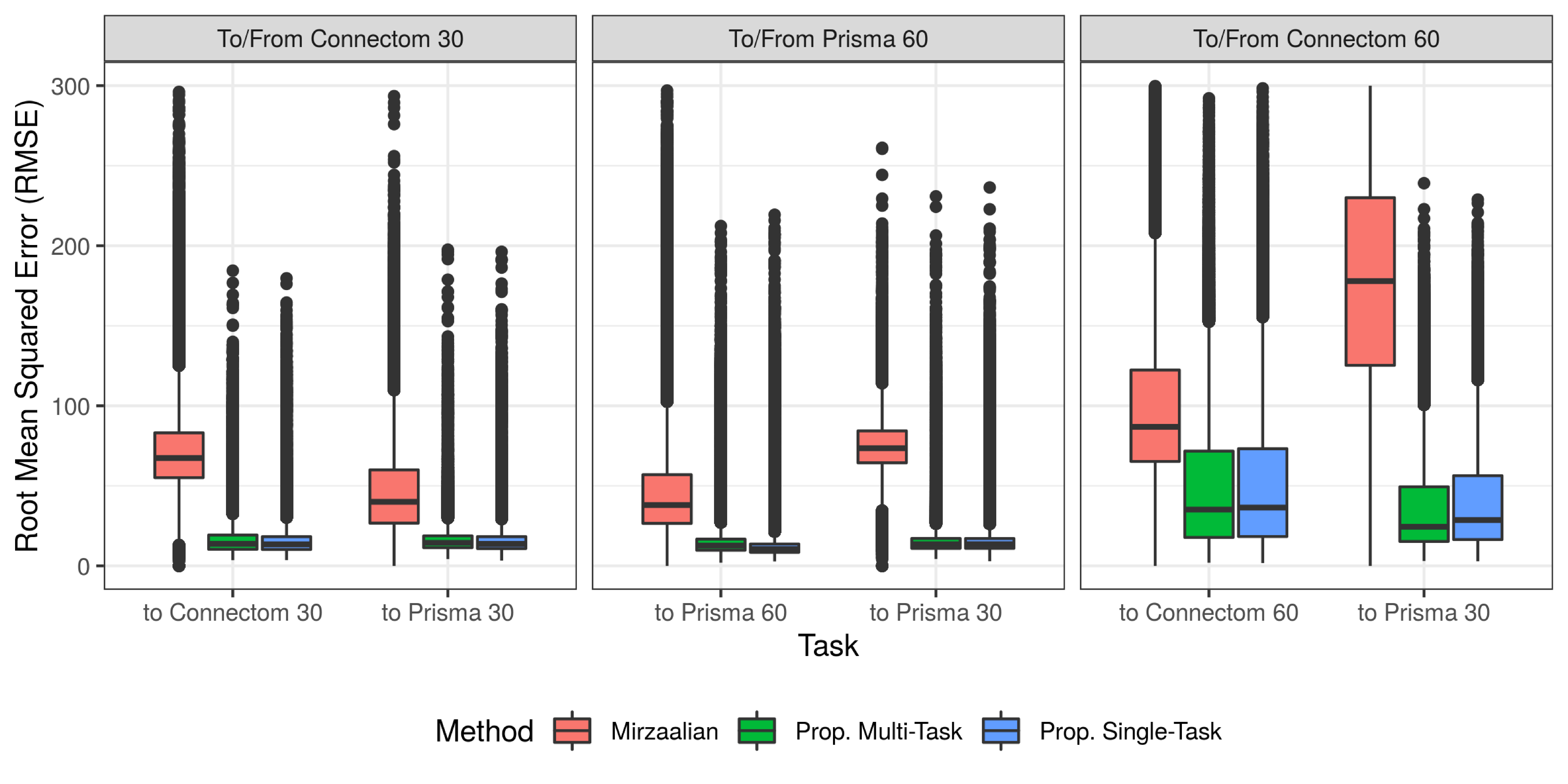}
\begin{tabular}{ | c | p{1.5cm} | p{2cm} | p{2cm} | p{3.2cm} |}
\multicolumn{1}{c}{} & \multicolumn{4}{ c }{ Post-hoc Adversarial Accuracy, Predicting $s$ from $z$ }\\ \cline{2-5}
\multicolumn{1}{c |}{} & Oracle & Full Model & No $\mathcal{L}_{marg}$  & No $\mathcal{L}_{marg}$ or $\mathcal{L}_{prior}$\\ \hline
Proposed Single-task, C30 
& 0.5 & 0.61 & 0.63 & 0.63
\\ \hline
Proposed Single-task, P60 
& 0.5 & 0.5 & 0.51 & 0.54
\\ \hline
Proposed Single-task, C60 
& 0.5 & 0.63 & 0.68 & 0.85
\\ \hline
Proposed Multi-task 
& 0.25  & 0.41 & 0.41 & 0.62 \\ \hline
\end{tabular}
\caption{At \textbf{top} we report the predictive accuracy for each of the methods, measured by RMSE per voxel in the original data domain, for each of the six tasks. 
At \textbf{bottom} we report the mean test accuracy for an adversary trained post-hoc to predict the site variable $s$ from encoding $z$, and adversarial accuracy for each ablation test.}
\label{fig:err}
\end{figure}

The authors split this dataset into 9 training subjects, 1 validation subject, and 5 held out-test subjects. One baseline comparison was found in the literature and compared to, Mirzaalian et al \cite{mirzaalian2018multi}, which relies on a template based solution. For the proposed method, a post-hoc adversary was fit to the learned code, as a lower-bound proxy \cite{NIPS2018_8122} for remaining mutual information $I(z,s)$; a further ablation test for each network component was conducted. As show in Figure \ref{fig:err}, the proposed methods significantly reduce predictive error, while also removing site information, as demonstrated by the post-hoc adversary.

\section{Discussion and Limitations}

Moyer et al. \cite{moyer2019scanner} presents an alternative method for harmonization, one using intermediate scanner invariant representations to remove scanner/site information. The proposed methods produces quality results, while removing site information from the intermediate encoding.

\midlacknowledgments{This summary was supported by NIH grants P41 EB015922, R01 MH116147, R56 AG058854, RF1 AG041915, and U54 EB020403, DARPA grant W911NF-16-1-0575, as well as the NSF Graduate Research Fellowship Program Grant Number DGE-1418060, and a GPU grant from NVidia.}

\bibliography{biblio}

\end{document}